\documentclass[aps,pre,showpacs,twocolumn]{revtex4}
\usepackage{graphicx}

\newcommand{\sn}{\,\mbox{sn}}

\begin{document}

\title{Analytic solutions for nonlinear waves in coupled reacting systems}

\date{\today}

\author{G. Abramson}
\altaffiliation[Permanent Address: ]{Centro At{\'o}mico Bariloche, Instituto
Balseiro and CONICET, 8400 S. C. de Bariloche, Argentina}
\email{abramson@cab.cnea.gov.ar}
\affiliation{Center for Advanced Studies and Department of Physics
and Astronomy, University of New Mexico, Albuquerque, New Mexico 87131}

\author{A. R. Bishop}
\affiliation{Theoretical Division and Center for Nonlinear Studies,
Los Alamos National Laboratory, Los Alamos, New Mexico 87545}

\author{V. M. Kenkre}
\affiliation{Center for Advanced Studies and Department of Physics
and Astronomy, University of New Mexico, Albuquerque, New Mexico 87131}

\begin{abstract}
We analyze a system of reacting elements harmonically coupled to nearest
neighbors in the continuum limit. An analytic solution is found for traveling
waves. The procedure is used to find oscillatory as well as solitary waves. A
comparison is made between exact solutions and solutions of the piecewise
linearized system, showing how the linearization affects the amplitude and
frequency of the solutions.
\end{abstract}

\pacs{05.45.-a, 05.45.Yv, 82.40.-g, 82.40.Ck}

\maketitle

\section{Introduction}

Consider a lattice consisting of a reaction system at each site $n$, described
by a site potential $V(u_n)$ with $u$ a one dimensional field. Neighboring
sites in the lattice interact through a potential $\phi(u)$. The equation of
motion:
\begin{equation}
\frac{d^2u_n}{dt^2}=-\phi'(u_n-u_{n-1}) + \phi'(u_{n+1}-u_n) - V'(u_n).
\label{lattice}
\end{equation}
describes, for instance, the time evolution of the amplitude of an oscillator,
the population of some ecological species, or the density of a chemical
component of the system. If $\phi$ is quadratic, the continuous form of
Eq.~(\ref{lattice}) is a wave equation augmented by an additive nonlinear term:
\begin{equation}
\frac{\partial^2 u}{\partial t^2}=
v^2 \,\frac{\partial^2 u}{\partial x^2} -V'(u),
\label{rw}
\end{equation}
where $v$ is the speed of linear waves in the absence of the nonlinear
potential $V$. The latter is determined by the details of the system under
consideration. A choice of wide applicability is the logistic reaction term.
Equation (\ref{rw}) is then:
\begin{equation}
\frac{\partial^2 u}{\partial t^2}=
v^2 \,\frac{\partial^2 u}{\partial x^2} +k\,u(1-u),
\label{rw0}
\end{equation}
whose interest resides in its relevance to chemical and population dynamics,
where the reaction term models an autocatalytical ``malthusian'' growth of $u$,
with a saturation due to intraspecies competition. We have taken rescaled
variables in Eq.~(\ref{rw0}) so that the carrying capacity of the medium is 1,
in order to simplify the notation.

Equation (\ref{rw0}) is very closely related to the so-called Fisher's
equation~\cite{murray} for reaction-diffusive systems, in which the left hand
side of Eq.~(\ref{rw0}) is replaced by the first-order time derivative
$\partial u/\partial t$. The two equations share the same nonlinear reaction
term but, respectively, represent opposite limits of completely coherent
(wave-like) and completely incoherent (diffusive) transport in the absence of
the reaction term. Generalizations of Fisher's equation that take into account
finite correlation time effects in the transport term have been studied
recently \cite{MHK,ABK}.

The logistic term $u(1-u)$ bears similarity to $\sin u$ between its zeros.
Replacement of the former term by the latter makes Eq.~(\ref{rw0}) the
sine-Gordon equation which describes, for example, the continuous limit of a
chain of harmonically coupled pendulums \cite{dodd}. The main difference
between the two equations is the periodicity of $\sin u$ and, consequently, an
infinite number of zeros (equilibria). The logistic term provides only two
equilibria. Obviously, we expect low amplitude solutions of Eq.~(\ref{rw0}) to
be qualitatively similar to sine-Gordon's solutions, and large amplitude
solutions (in particular solitary wave solutions connecting the two equilibria)
to be quite different from the solitonic solutions of the sine-Gordon equation.

We analyze in this paper \emph{traveling wave} solutions of Eqs.~(\ref{rw0}),
which can be calculated exactly by analytical means. It is customary to analyze
a piecewise linear simplification of a nonlinear system like Eq.~(\ref{rw0}),
such as is done in Ref.~\cite{MHK}. Our analytic solution will provide us an
opportunity to assess the validity of a piecewise linearization of the problem.
Traveling waves of a related equation are analyzed in Section~\ref{kinks}.

\section{Traveling waves}

We look for traveling-wave solutions of Eq.~(\ref{rw0}), moving in the
direction of increasing $x$, by means of the ansatz: $u(x-ct)=U(z)$. In general
$c$, the speed of the nonlinear waves, will be different from $v$, the speed of
the linear waves dictated by the medium. The ordinary differential equation
describing the shape of the wave is:
\begin{equation}
(c^2-v^2)\,U'' = k f(U) = -\,\frac{\partial V}{\partial U}.
\label{oscillator}
\end{equation}

A mechanical analogy involving a particle of mass $m=|c^2-v^2|$ is immediately
available (see e.g.~\cite{MHK}). Explicit solutions can be found by multiplying
Eq.~(\ref{oscillator}) by $U'$ and integrating to find:
\begin{equation}
\frac{1}{2}(c^2-v^2)U'^2-E=\int k f(U) dU \equiv -V(U),
\end{equation}
where $E$ is the ``energy'' of the particle, that depends on the initial
conditions. Integrating once more:
\begin{equation}
z-z_0=\int_{U_0}^U \frac{dU'}{\sqrt{\frac{2}{c^2-v^2}[E-V(U')]}}.
\label{waves}
\end{equation}

This is an elliptic integral that can be solved exactly since the radicand is a
cubic function of $U$. This is done in Section~\ref{exact}.

\begin{figure}[h]
\centering
\resizebox{\columnwidth}{!}{\includegraphics{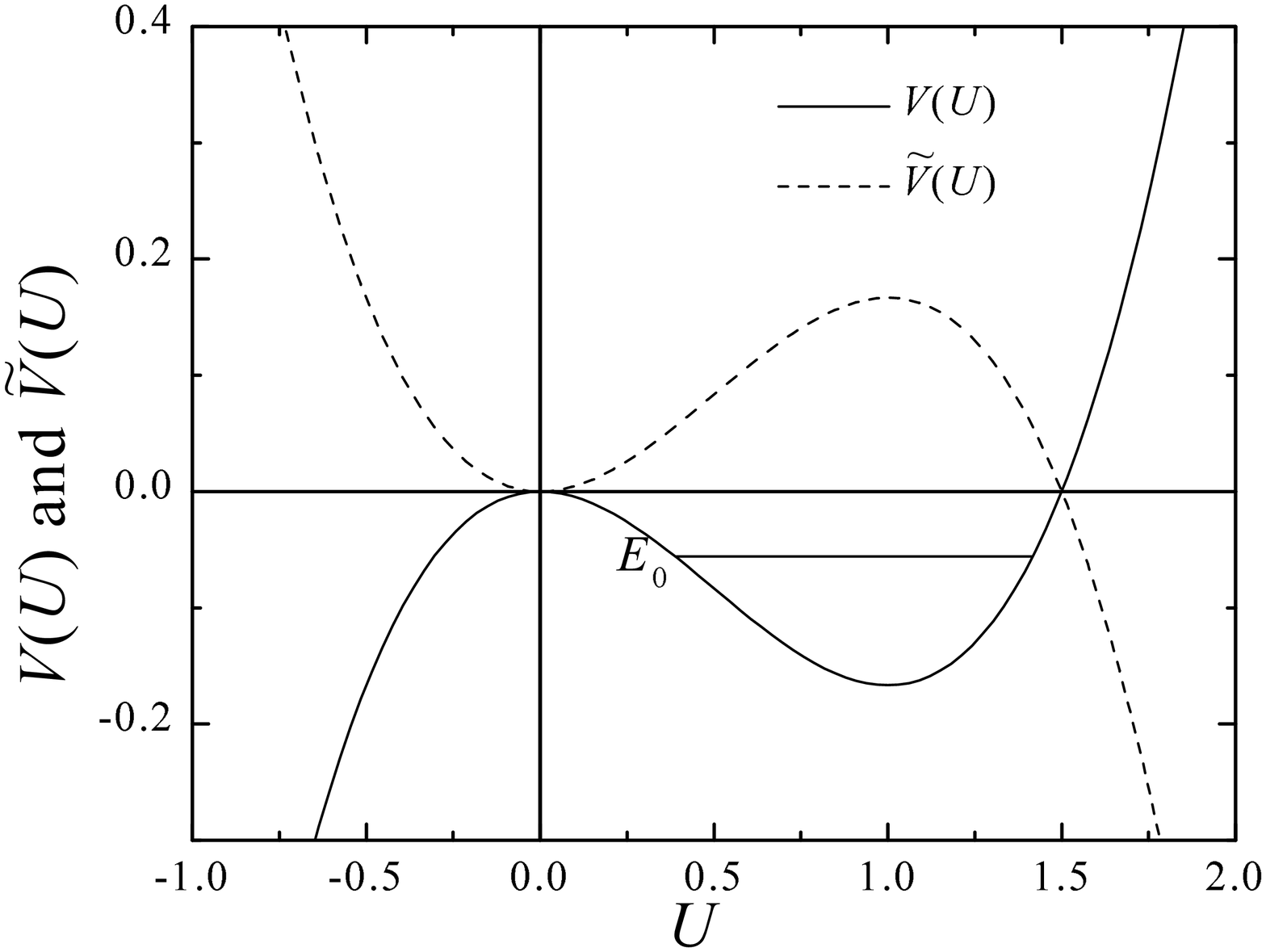}}
\caption{The potentials of the mechanical analog of Eq.~(\protect\ref{oscillator}).
The cases shown are $V$ and $\tilde V$, corresponding to faster than $v$ and
slower than $v$ traveling waves respectively.}
\label{potentials1}
\end{figure}

The potential of this mechanical system is $V(U)=k(U^3/3-U^2/2)$ if $c>v$, and
$\tilde{V}(U)=-V(U)$ if $c<v$ (refer to Fig. \ref{potentials1} for a
comparison). The potential $V$ has a maximum $V=0$ at $U=0$ and a minimum at
$U=1$. Thus, if $c>v$, the system performs oscillations around $U=1$, with the
amplitude $U$ always remaining positive, provided that the energy satisfies
$-k/6<E<0$. If $E=0$ there is a solitary-wave solution with the shape of a
traveling pulse, growing from $U=0$ to $U=1.5$ and back to $U=0$.

The potential $\tilde{V}$, instead, has a minimum at $U=0$ and a maximum at
$U=1$. The system oscillates around $U=0$ if the energy is positive and smaller
than a finite value ($0<E<k/6$), and a traveling trough if it is exactly this
value. Since we are interested in situations where $u$ describes a density or a
population, we will exclude these solutions that contain a negative amplitude
in the present discussion. Nevertheless, they can be analyzed without
difficulty in the same way.

\subsection{General solutions}
\label{exact}

As we noted, $U(z)$ can be found exactly by performing the elliptic integral in
Eq.~(\ref{waves}). Note the following simplified form \cite{mathematica}:
\begin{equation}
\int\frac{dU}{\sqrt{1-U^3+U^2}}=A
{\cal F}\left(\arg\sin\sqrt{B+CU}|\gamma\right),
\label{elliptic}
\end{equation}
where ${\cal F}(\phi|\gamma)$ is the elliptic integral of the first kind, and
$A$, $B$, $C$, and $\gamma$ are constants involving the roots of $1-U^3+U^2$.
The constant $\gamma$ is termed the \emph{modulus} of the elliptic integral,
not to be confused with the \emph{parameter} which is $m=\gamma^2$. (We are
following the nomenclature of Ref.~\cite{byrd}, chapter 8.1, but denoting the
modulus $\gamma$ instead of $k$ to avoid confusion with the reaction constant.)
The relation between the elliptic integral ${\cal F}$ and the jacobian elliptic
function $\sn (z|\gamma)$ suggests that the solution of Eq.~(\ref{oscillator})
is of the form:
\begin{equation}
U(z)=A+B\sn ^2(\beta z|\gamma),
\label{sol}
\end{equation}
where $A$, $B$, $\beta$ and $\gamma$ are constants, depending on the initial
conditions, that remain to be determined. Differentiating Eq.~(\ref{sol}) with
respect to $z$, squaring and multiplying by $m/2$, we arrive at:
\begin{eqnarray}
\frac{m}{2} U'^2&=& -2m\beta B(U-A) \nonumber \\*
& &\times\left[1-\frac{1-\gamma}{B}(U-A)+\frac{\gamma}{B^2}(U-A)^2\right].
\label{energy}
\end{eqnarray}
Equation~(\ref{energy}) is the kinetic energy, that we can equate to $E-V(U)$,
which in turn we can write in the form:
\begin{eqnarray}
E-V(U)&=& E-\frac{k}{3}\,U^3+\frac{k}{2}\,U^2 \nonumber \\*
&=&\frac{k}{3}\,(U-U_1)(U-U_2)(U-U_3),
\label{energy2}
\end{eqnarray}
where $U_1$, $U_2$ and $U_3$ are the zeroes of $E-V$, and depend on the initial
conditions. By equating Eq.~(\ref{energy2}) and Eq.~(\ref{energy}) we can
express the unknowns $A$, $B$, $\beta$ and $\gamma$ as functions of $U_1$,
$U_2$, $U_3$ and $k$. Because these expressions are intricate and do not add
much to the analysis, we do not display them. However, one of the consequences
of these expressions is the following dispersion relation, relating the speed
of the nonlinear waves to system parameters:
\begin{equation}
c=\left[
        v^2+\frac{k}{12\beta}\left(\alpha+\sqrt{2\alpha^2-(U_1-U_3)^2}\right)
  \right]^{1/2},
\end{equation}
where $\alpha=-U_1+2U_2-U_3$.

\begin{figure}
\centering
\resizebox{\columnwidth}{!}{\includegraphics{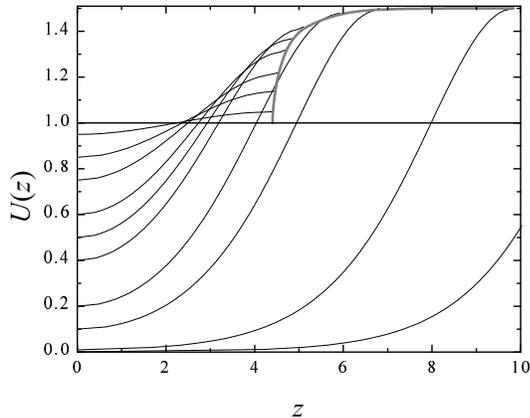}}
\caption{Half waves of the periodic solutions of different energy. Greater
amplitude (and lower frequency) correspond to higher energy. The parameters are
$k=1$, $m=2$. }
\label{halfwaves}
\end{figure}

Equation (\ref{waves}) can also be integrated numerically to yield $z(U)$ for a
given value of the energy $E$. In Fig.~\ref{halfwaves} we show half-waves for
several energies. The complete solutions are periodic functions of $z$,
symmetric around their maxima, and we are plotting here only one half of a
period. These solutions correspond to waves of progressively lower energy,
smaller amplitude and, as can be seen, corresponding higher frequency. A grey
line relating amplitude to period connects the half periods, and can be seen to
converge to a value of the half period of $\sqrt{2}\pi\approx 4.44$,
corresponding to the harmonic small oscillations. Waves of diverging period
eventually become a traveling pulse, that we describe below.

\subsection{Validity of piecewise linearization procedures}

A piecewise linearization of Eq.~(\ref{rw0}) can be made, as in related models
of reaction-diffusion processes (see for example \cite{mckean,koga,ohta} for
excitable systems, \cite{schat} for an electrothermal instability, and the
generalization of Fisher's equation found in \cite{MHK}). Following \cite{MHK},
the potential of the piecewise linear oscillator can be written as:
\begin{equation}
V(U)=\left\{\begin{array}{cl}
            \displaystyle -V_0k U^2 /(2a),&\;\;\; U\leq a \\
            \displaystyle V_0\frac{k}{b-a}(U^2/2-bU),&\;\;\; U\geq a.
            \end{array}
     \right.
\end{equation}
with free parameters, $a$, $b$ and $V_0$, that can be adjusted to match the
nonlinear one. The corresponding oscillator system is:
\begin{eqnarray}
mU''&=&\frac{k}{ma}U\;\;\;\;\;\; (U\leq a), \label{harmonic1}\\
mU''&=&-\frac{k}{m(b-a)}(U-b)\;\;\;\; (U\geq a),
\label{harmonic2}
\end{eqnarray}
whose solution can be found explicitly. Equations~(\ref{harmonic1}) and
(\ref{harmonic2}) describe harmonic oscillators, the first one with a positive
restitutive force, so that the solution will be unstable (if $m>0$). If the
initial condition $U_0$ is lower than $a$ and has zero velocity, the general
solution is a smooth match (a continuous match of $U$ and $U'$) of the
solutions of both equations (\ref{harmonic1},\ref{harmonic2}). If $U_0>a$ the
solution is just that of (\ref{harmonic2}).

We define natural frequencies: $\omega_1=\sqrt{k/m}$ and
$\omega_2=\sqrt{k/[m(b-a)]}$. The solution to (\ref{harmonic1},\ref{harmonic2})
is:
\begin{equation}
U(z)=\left\{\begin{array}{ll}
            \displaystyle U_0 \cosh (\omega_1 z),&\;\;\; U\leq a, \\
            \displaystyle A\cos(\omega_2 z)+B \sin(\omega_2 z) +b,&\;\;\; U\geq a.
            \end{array}
     \right.
\end{equation}
where $A$ and $B$ are found as a solution to the following linear system:
\begin{equation}
\begin{array}{rcrcl}
\cos(\omega_2 z_a)\,A &+& \sin(\omega_2 z_a)\,B &=& a-b, \\
-\omega_2 \sin(\omega_2 z_a)\,A &+&\omega_2 \cos(\omega_2 z_a)\,B &=&
U_0\omega_1\sinh(\omega_1 z_a),\\
\end{array}
\end{equation}
where $z_a$ is the time it takes the unstable solution to grow from $U_0$ to
$a$, namely: $z_a=\omega_1^{-1}\arg \cosh(a/U_0)$.

When the initial condition is greater than $a$, the solution is:
\begin{equation}
U(z)=(U_0 -b) \cos(\omega_2 z) +b,\;\;\;\;\;(U_0\geq a).
\end{equation}

The two kinds of solutions (those of the nonlinear system and those of the
piecewise linear) have the same qualitative shape. The piecewise linearization
of the system is harmonic around the minimum at $U=b$, so there are harmonic
oscillations of amplitude up to $b-a$ (oscillations of greater amplitude feel
the nonlinearity at $U=a$). The frequency of these oscillations is
$\omega_{lin}=\sqrt{k/[m(b-a)]}$. If $b=1$, $a=1/2$ and $V_0=1/4$ (to match the
logistic case $U(1-U)$ as well as possible) then $\omega_{lin}=\sqrt{k/2m}$
which is $\sqrt 2$ smaller than what is found for the nonlinear system.

\begin{figure}
\centering
\resizebox{\columnwidth}{!}{\includegraphics{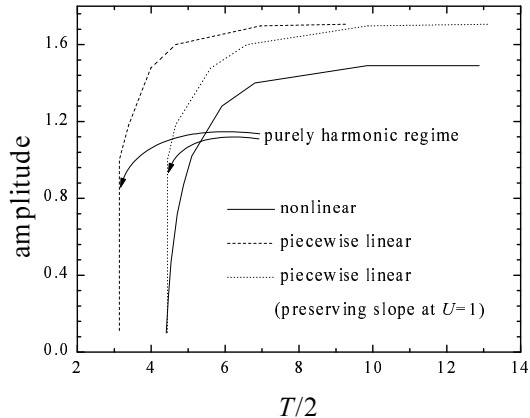}}
\caption{Amplitude vs. half period of traveling wave solutions. Results are shown
for the fully nonlinear system and for two piecewise linearizations: one that
preserves the value $f(a)$ and one that preserves the slope of $f$ at $u=b$, as
shown in the legend. The parameters are: $k=1$, $m=2$, $a=1/2$, $b=1$.}
\label{amplitude}
\end{figure}

This discrepancy can be seen in Fig.~\ref{amplitude} where the relationship
between the amplitude and the half period is shown, in a way similar to a
dispersion relation. The nonlinear solution is shown as the full line. The
piecewise linear solution that preserves the intensity of the nonlinearity is
shown as the dashed line. It's frequency at small amplitude overestimates the
nonlinear one by a factor of $\sqrt 2$.

\begin{figure}
\centering
\resizebox{\columnwidth}{!}{\includegraphics{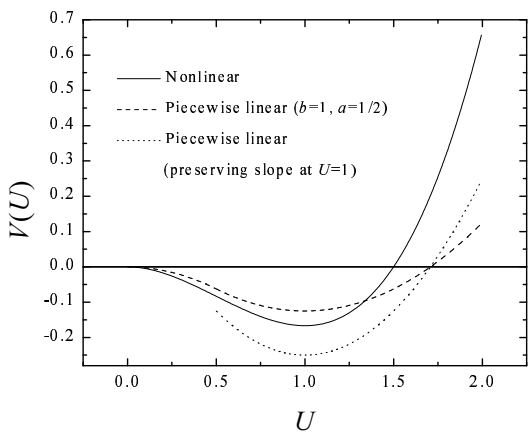}}
\caption{The potential of the mechanical analog of the fully nonlinear and the
piecewise linear systems. The case shown corresponds to $c>v$.}
\label{potentials}
\end{figure}

Of course, the piecewise linear procedure can be adjusted to fit the parameter
we want, such as the slope at $U=1$. In this case the frequency of oscillations
is $\sqrt{k/m}$, but $f(a=1/2)=1/2$ instead of 1/4. This case is shown in
Fig.~\ref{amplitude} as the dotted line; it has the correct frequency at small
amplitude. Both linearizations overestimate the amplitude of the nonlinear
waves. Fig. \ref{potentials} shows the three potentials. In can be seen that
the amplitude of traveling waves of a certain energy will always be
overestimated in the piecewise linear system.

\subsection{Traveling pulse at $E=0$}

When $E=0$ (and $c>v$) the solution is a traveling solitary pulse instead of a
periodic wave. It corresponds to the case $\gamma=1$, when
$\sn(\phi|1)=\tanh(\phi)$, and it represents the trajectory of the particle
leaving the maximum of $V$ at $U=0$ towards the right (see
Fig.~\ref{potentials}) and bouncing back to $U=0$ after reaching the maximum
amplitude $U=1.5$. In this case, the integral (\ref{waves}) simplifies
considerably, and the pulse can be written down explicitly as:
\begin{equation}
U(z)=\frac{3}{2}\left\{
     1-\left[
       \tanh\left(
         -\frac{z}{2}\sqrt\frac{k}{c^2-v^2}
         \right)
       \right]^2
     \right\}.
\label{pulse}
\end{equation}

The speed of the pulse is related to its width. From the second moment of
(\ref{pulse}) we find the following expression for the velocity as a function
of the parameters:
\begin{equation}
c(\sigma)=\sqrt{v^2+k\left(\sqrt{2}\pi^2\sigma^2\right)^{2/3}},
\end{equation}
where $\sigma$ is the variance of (\ref{pulse}). It can be seen that wider
pulses move faster than narrow ones, much in the way that the less steep fronts
are faster in Fisher's equation \cite{murray}.

This solitary wave solution is different from the kink-type soliton of the
sine-Gordon equation, despite the similarity between the two equations. The
potential in the sine-Gordon case is sinusoidal, allowing the state of the
field $u$ to shift by amounts of $2\pi$ in a kink. In the present case the
potential $V(u)$ continues growing as $u\rightarrow\infty$, and the state $u$
is forced back to $u=0$ after the excursion represented by the pulse
(\ref{pulse}).

\section{Front solutions for a related system}
\label{kinks}

It is possible to approach our model of coupled reactors from another
direction. Suppose that at each site of the lattice we have a system described
by a first order equation of chemical or population dynamics:
\begin{equation}
\frac{du_n}{dt}=f(u_n).
\label{logistic}
\end{equation}
So, if $f$ is logistic, the ``force'' (deriving (\ref{logistic}) with respect
to time) is the cubic $f'(u)f(u)=k(1-2u)u(1-u)$ and the corresponding
generalization of the wave equation is:
\begin{equation}
\frac{\partial^2 u}{\partial t^2}=
v^2 \,\frac{\partial^2 u}{\partial x^2} +ku(1-u)(1-2u).
\label{rw1}
\end{equation}
Equation (\ref{rw1}) is very similar to the $\phi^4$ equation of particle
physics \cite{dodd}. But in Eq.~(\ref{rw1}) the $u^4$ potential is inverted and
it has as a consequence only one stable vacuum state instead of two.

Solutions of Eq.~(\ref{rw1}) are qualitatively similar to those of
Eq.~(\ref{rw0}) around and near the equilibrium $u=0$. Solutions far from the
equilibrium can be analyzed with the same methods as those of Eq.~(\ref{rw0}),
since the integrals involved in (\ref{waves}) are also elliptic. In fact,
Eq.~(\ref{rw1}) is similar to a Klein-Gordon $\phi^4$ model, but with an
inverted potential. The solitary wave in this case is of the kink type, and it
is straightforward to calculate its shape as $z(U)$:
\begin{equation}
z(U)=\sqrt{
           \frac{m}{2k}
          }
          \frac{ U (1-U) \log{[(1-U)/U]}
               }
               {
                \sqrt{U^4/2-U^3+U^2/2}
               }.
\label{kink}
\end{equation}
This solution is a traveling front, connecting the hyperbolic points located at
the two maxima of the potential, at $U=0$ and $U=1$. It represents the invasion
of one of the states by the other. A kink from $U=0$ to $U=1$ as well as an
``anti-kink'' from $U=1$ to $U=0$ are possible, opening interesting
possibilities of multiple interacting kinks, which will be analyzed elsewhere.

\begin{figure}
\centering
\resizebox{\columnwidth}{!}{\includegraphics{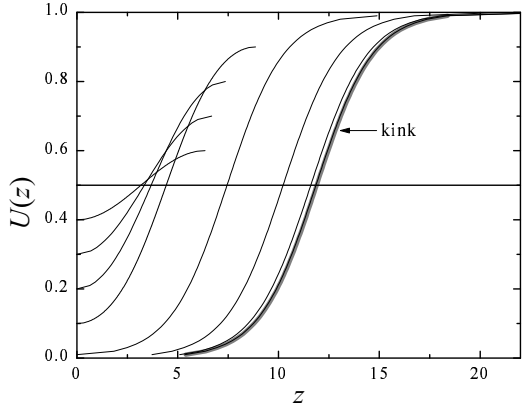}}
\caption{Half waves of the periodic solutions of Eq.~(\protect\ref{rw1}) (thin
black lines) and kink solution (\protect\ref{kink}) (thick grey line).
Different lines correspond to different negative energies of the mechanical
analog. The kink is the solution with zero energy.}
\label{kinkfig}
\end{figure}

Oscillatory wave solutions are plotted in Fig.~\ref{kinkfig} for several
initial conditions. Lower energies tend to accumulate to the kink solution
(\ref{kink}), shown with a thicker line in grey.

\section{Conclusion}

We have analyzed two systems of reacting elements harmonically coupled to first
neighbors. The equation of motion in the continuous limit is, in each case, a
wave equation with a nonlinear reaction term. The transport character in the
systems is, thus, fully coherent, in contrast to a case such as that of
Fisher's equation~\cite{murray}, in which it is fully incoherent. Intermediate
transport coherence has been analyzed recently \cite{MHK,ABK} through the
incorporation of memory functions and it has been shown how the incoherent
(Fisher) limit may be obtained for a memory with infinitely fast decay. The
opposite limit wherein the memory is constant is Eq.~(\ref{rw0}) in the present
paper, while a related system is Eq.~(\ref{rw1}). The latter represents a set
of logistic systems, coupled harmonically. There is an interesting difference
between its solutions and those of other extended reaction models. In Fisher's
reaction-diffusion equation, traveling front solutions of a single kind are
found. In the generalized system studied by us elsewhere~\cite{ABK}, two kinds
of fronts---a front and an ``anti-front''---have been found, which however
cannot coexist: each exists for a different set of system parameters. By
contrast, in the purely wave-like system we have studied in the present paper,
the coexistence of kinks and anti-kinks is certainly possible. This opens an
interesting possibility of multiple interacting fronts.

\begin{acknowledgments}
This work was supported in part by the Los Alamos National Laboratory via a
grant made to the University of New Mexico (Consortium of the Americas for
Interdisciplinary Science) and by the National Science Foundation's Division of
Materials Research via grant No. DMR0097204. G. A. thanks the support of the
Consortium of the Americas for Interdisciplinary Science and the hospitality of
the University of New Mexico.
\end{acknowledgments}


\begin{thebibliography}{99}

\bibitem{murray} J. D. Murray, \emph{Mathematical Biology}, 2nd ed. (Springer,
New York, 1993).

\bibitem{MHK} K. K. Manne, A. J. Hurd and V. M. Kenkre, Phys. Rev. E
\textbf{61}, 4177 (2000).

\bibitem{ABK} G. Abramson, A. R. Bishop and V. M. Kenkre, preprint
nlin.PS/0107043.

\bibitem{dodd} R. K. Dodd \emph{et al.}, \emph{Solitons and Nonlinear Wave
Equations} (Academic Press, London, 1982).

\bibitem{mathematica} Wolfram Research, Inc., \emph{Mathematica}, version 4,
(Wolfram Research, Inc., Champaign, IL, 1999).

\bibitem{byrd} P. F. Byrd and M. D. Friedman, \emph{Handbook of elliptic
integrals for engineers and scientists}, 2nd ed., (Springer-Verlag, Berlin, New
York, 1971).

\bibitem{mckean} H. P. McKean, Adv. Math. \textbf{4}, 209 (1970).

\bibitem{koga} S. Koga and Y. Kuramoto, Progr. Theor. Phys. \textbf{63}, 106
(1980).

\bibitem{ohta} T. Ohta, A. Ito and A. Tetsuke, Phys. Rev. A \textbf{42}, 3225
(1990); T. Ohta, Progr. Theor. Phys. \textbf{S. 99}, 425 (1989); T. Ohta, M.
Mimura and R. Kobayashi, Physica D \textbf{34}, 115 (1989).

\bibitem{schat} C. Schat and H. S. Wio, Physica A \textbf{180}, 295 (1992).

\end{thebibliography}
\end{document}